# Non-blind watermarking of network flows

Amir Houmansadr*, Negar Kiyavash, and Nikita Borisov

**Abstract**—Linking network flows is an important problem in intrusion detection as well as anonymity. Passive traffic analysis can link flows but requires long periods of observation to reduce errors. Active traffic analysis, also known as flow watermarking, allows for better precision and is more scalable. Previous flow watermarks introduce significant delays to the traffic flow as a side effect of using a blind detection scheme; this enables attacks that detect and remove the watermark, while at the same time slowing down legitimate traffic. We propose the first non-blind approach for flow watermarking, called RAINBOW, that improves watermark invisibility by inserting delays hundreds of times smaller than previous blind watermarks, hence reduces the watermark interference on network flows. We derive and analyze the optimum detectors for RAINBOW as well as the passive traffic analysis under different traffic models by using hypothesis testing. Comparing the detection performance of RAINBOW and the passive approach we observe that both RAINBOW and passive traffic analysis perform similarly good in the case of uncorrelated traffic, however, the RAINBOW detector drastically outperforms the optimum passive detector in the case of correlated network flows. This justifies the use of non-blind watermarks over passive traffic analysis even though both approaches have similar scalability constraints. We confirm our analysis by simulating the detectors and testing them against large traces of real network flows.

**Index Terms**—Traffic analysis, flow watermarking, non-blind watermarking, hypothesis testing.

## 1 INTRODUCTION

Internet attackers commonly relay their traffic through a number of (usually compromised) hosts in order to hide their identity. Detecting such hosts, called stepping stones, is therefore an important problem in computer security. The detection proceeds by finding correlated flows entering and leaving the network. Traditional approaches have used patterns inherent in traffic flows, such as packet timings, sizes, and counts, to link an incoming flow to an outgoing one [1], [2], [3], [4], [5]. More recently, an active approach called *watermarking* has been considered [6], [7]. In this approach, traffic characteristics of an incoming flow are actively perturbed as they traverse some router to create a distinct pattern, which can later be recognized in outgoing flows. These techniques also have relevance to anonymous communication, as linking two flows can be used to break anonymity, and both passive traffic analysis [8], [9] and active watermarking [10], [11], [12] have been studied in that domain as well.

The choice between passive and active techniques for traffic analysis exhibits a tradeoff. Passive approaches require observing relatively long-lived network flows, and storing or transmitting large amounts of traffic characteristics. Watermarking approaches are more efficient, with shorter observation periods necessary. They are also *blind*: rather than storing or communicating traffic patterns, all the necessary information is embedded in the flow itself. This, however, comes at a cost: to ensure robustness, the watermarks introduce large delays (hundreds of milliseconds) to the flows, interfering with the activity of benign users, and making them subject to attacks [13], [14].

Motivated by this, we propose a new category for network flow watermarks, the *non-blind flow watermarks*. Non-blind watermarking lies in the middle of passive techniques and (blind) watermarking techniques: similar to passive techniques (and unlike blind watermarks), non-blind watermarks will record traffic pattern of incoming flows and correlate them with outgoing flows. On the other side, similar to blind watermarks (and unlike passive techniques), non-blind watermarking aids traffic analysis by applying some modifications to the communication patterns of the intercepted flows. We develop and prototype the first non-blind flow watermark, called RAINBOW. RAINBOW records the *timing* pattern of incoming flows and correlate them with the timing pattern of the outgoing flows. On each incoming flow, RAINBOW also inserts a watermark by delaying some packets, after recording the received timings. As such a watermark is generated independently of the flows, this will diminish the effect of natural similarities between two unrelated flows, and allow a flow linking decision to be made over a much shorter time period. RAINBOW uses spread-spectrum techniques to make the delays much smaller than previous work. RAINBOW uses delays that are on the order of only a few milliseconds; this means that RAINBOW watermarks not only do not interfere with traffic patterns of normal users, they are also virtually *invisible*, since the delays are of the same magnitude as natural network jitter. In [15] we use different information theoretical tools to verify the invisibility of RAINBOW, and demonstrate its high per-

• A. Houmansadr, N. Kiyavash, and N. Borisov are with the University of Illinois at Urbana-Champaign (UIUC), Urbana, IL.
Address: 342 Coordinated Science Laboratory, 1308 West Main St., Urbana, IL 61801
Phone: +1 (217) 722-1761, Fax: +1 (217) 244-5685
E-mail: {ahouman2,kiyavash,nikita}@illinois.edu

formance in linking network flows through a prototype implementation over the PlanetLab [16] infrastructure.

In this paper, we thoroughly analyze the detection performance of RAINBOW non-blind watermark, and compare it with that of passive traffic analysis schemes. By using *hypothesis testing* mechanisms from the detection and estimation theory [17], we find the optimum detection schemes for RAINBOW as well as the optimum passive detectors under different models for network traffic. Modeling real-world network traffic is a complicated problem as it depends on many different parameters; as a result, we only consider two extreme models of the network traffic: (1) independent flows where each flow is modeled as a Poisson process (traffic model A), and, (2) completely correlated flows where all flows are considered to have similar timing patterns (traffic model B). We assume that any real-world traffic model lies in the middle of these two extreme models. Our analysis leads to the following important conclusions:

i) Non-blind watermarking *always* performs a better detection than passive traffic analysis. This is an essential result in motivating the use of non-blind watermarks over passive traffic analysis, since both have similar scalability constraints, i.e., both approaches have $O(n)$ communication overheads and $O(n^2)$ computation overheads [15]. Not that this point is not necessary (nor is always true) to motivate the use of traditional (blind) watermarks over passive traffic analysis, since blind watermarks provide much better scalability (i.e., $O(1)$ communication overhead and $O(n)$ computation overhead [15] ).

ii) Our analysis shows that the performance advantage of non-blind watermarking (over passive schemes) is only marginal for uncorrelated network traffic, while it is very significant for correlated network traffic. This knowledge can be used to decide the best traffic analysis approach in various applications. We validate our analysis through simulating the detection schemes on real network traces. In particular, we show that for highly correlated traffic, e.g., same webpage downloads, passive traffic analysis performs very poorly while a RAINBOW watermark is highly effective.

iii) We also show (through both analysis and experiments) that the optimum watermark detector derived for correlated traffic (namely $SLCorr$) also performs very good for uncorrelated traffic (while the optimum watermark detector for uncorrelated traffic does not do well for correlated traffic). This allows one to use $SLCorr$ as the sole watermark detector regardless of the type of traffic being observed. This is especially useful in real-world applications where the observed traffic is a mixture of different flow types.

Note that in this paper we do not discuss the performance advantage of non-blind watermarks over traditional blind watermarks, as this has been justified in [15].

The rest of this paper is organized as follows: we review the problem of stepping stone detection and existing schemes in Section 2. Our RAINBOW scheme is presented in Section 3. In Section 4, we use hypothesis testing to find and analyze the optimum likelihood ratio detectors for passive and non-blind active (watermark) approaches under different traffic models, and analyze their false error rates. In Section 5, we validate the analysis results through simulation of the detection schemes over real network traces. Finally, the paper is concluded in Section 6.

## 2 BACKGROUND

In this section, we review the problem of detecting stepping stones and then review both the passive and active approaches to the problem. We compare the advantages and disadvantages of the two techniques, motivating our approach.

### 2.1 Stepping Stone Detection

A stepping stone is a host that is used to relay traffic through an enterprise network to another remote destination. Stepping stones are used to disguise the true origin of an attack. Detecting stepping stones can help trace attacks back to their true source. Also, stepping stones are often indicative of a compromised machine. Thus detecting stepping stones is a useful part of enterprise security monitoring.

Generally, stepping stones are detected by noticing that an outgoing flow from an enterprise matches an incoming flow. Since the relayed connections are often encrypted (using SSH [18], for example), only characteristics such as packet sizes, counts, and timings are available for such detection. And even these are not perfectly replicated from an incoming flow to an outgoing flow, as they are changed by padding schemes, retransmissions, and jitter. As a result, statistical methods are used to detect correlations among the incoming and outgoing flows. We next review the passive and active approaches.

### 2.2 Passive Traffic Analysis

In general, passive traffic analysis techniques operate by recording characteristics of incoming streams and then correlating them with outgoing ones. The right place to do this is often at the border router of an enterprise, so the overhead of this technique is the space used to store the stream characteristics long enough to check against correlated relayed streams, and the CPU time needed to perform the correlations. In a complex enterprise with many interconnected networks, a connection relayed through a stepping stone may enter and leave the enterprise through different points; in such cases, there is additional communications overhead for transmitting traffic statistics between border routers.

The passive schemes have explored using various characteristics for correlating streams. Zhang and Paxson [2] model interactive flows as on–off processes and detect linked flows by matching up their on–off behavior.

Wang et al. [4] focus on inter-packet delays, and consider several different metrics for correlation. More recently, He and Tong used packet counts for stepping stone detection [19].

Donoho et al. were the first to consider intruder evasion techniques [3]. They defined a *maximum-tolerable-delay* (MTD) model of attacker evasion and suggested wavelet methods to detect stepping stones while being robust to adversarial action. Blum et al. used a Poisson model of flows to create a technique with provable upper bounds on false positive rates [5], given the MTD model. However, for realistic settings, their techniques require thousands of packets to be observed to achieve reasonable rates of false errors.

### 2.3 Watermarks

To address some of the efficiency concerns of passive traffic analysis, Wang et al. proposed the use of watermarks [6]. In this scenario, a border router will modify the traffic timings of the incoming flows to contain a particular pattern—the watermark. If the same pattern is present in an outgoing flow, a stepping stone is detected.

Watermarks improve upon passive traffic analysis in two ways. First, by inserting a pattern that is uncorrelated with any other flows, they can improve the detection efficiency, requiring smaller numbers of packets to be observed (hundreds instead of thousands) and providing lower false-positive rates ($10^{-4}$ or lower, as compared to $10^{-2}$ with passive watermarks). Second, they can operate in a *blind* fashion: after an incoming flow is watermarked, there is no need to record or communicate the flow characteristics, since the presence of a watermark can be detected independently. The detection is also potentially faster, as here is no need to compare each outgoing flow to all the incoming flows within the same time frame.

Watermarking techniques for network flows have been based on existing techniques for multi-media watermarking. For example, Wang et al. based their scheme on QIM watermarks [20]. Two other watermark schemes [7], [11] are based on patchwork watermarking [21], and Yu et al. [12] developed one based on spread-spectrum techniques [22]. Some of the schemes target anonymous communication rather than stepping stones as the application area (both involve the problem of linking flows), but the techniques for both are comparable.

### 2.4 Watermark Properties

To motivate our design, we first propose some desirable properties of network flow watermarks. First of all, a watermark should be *robust* to modifications of the traffic characteristics that will occur inside an enterprise network, such as jitter. Watermarks should also be resilient to an adversary who actively tries to remove them from the flow, a property we call *active robustness*. The watermarks should also introduce little *distortion*, in that they should not significantly impact the performance of the flows. This is important because in a stepping-stone scenario, most watermarked flows will be benign. Finally, watermarks should be *invisible* even to attackers who specifically try to test for their presence.

Looking at previous designs, all of them fail to be invisible: the watermarks introduce large delays, on the order of hundreds of milliseconds, on some packets, which can be easily detected by an attacker [13]. In fact, they cannot even be considered low-distortion, as such large delays are easily noticeable and bothersome to legitimate users. The watermarks are also not actively robust, as demonstrated by recent attacks [13], [14].

We also observe that active robustness and invisibility are likely to be impossible to achieve at the same time. This is because to be invisible, the watermark can only introduce minute changes to the packet stream. In particular, it cannot introduce jitter of more than a few milliseconds, since otherwise it will be possible to tell it apart from the natural network jitter. However, an active attacker will be willing to introduce large delays to the network; for example, the maximum tolerable delay suggested in previous work is 500ms. As such, he will be able to destroy any low-order effects that will be introduced by the watermark.

Further, it is easy to imagine an attacker determined to hide his tracks using even more drastic measures, such as using dummy packets to generate a completely independent Poisson process [5], which will render any linking techniques ineffective. As such, we decided to design a watermark scheme that is robust to normal network interference, though not actively robust, and is invisible. This will serve to detect stepping stones where attackers are unwilling (or unable) to actively distort their stream as it crosses a stepping stone. Further, as the watermark will be invisible, attackers will not be able to tell if they are being traced and thus will be less likely to try to apply costly watermark countermeasures.

## 3 RAINBOW WATERMARK

We next present the design of a new watermark scheme we call RAINBOW, for Robust And Invisible Non-Blind Watermark. Our scheme is robust (to passive interference) and invisible. However, to achieve invisibility while maintaining detection efficiency, we make the scheme *non-blind*; that is, incoming flows timings are recorded and compared with the timings of outgoing flows. This allows us to make a robust watermark test with even low-amplitude watermarks.

The RAINBOW watermark embedding process is shown in Figure 1. Suppose that a flow with the packet timing information $\{t_i^u | i = 1, ..., n + 1\}$ enters border router where it is to be watermarked (we use the superscript $u$ to denote an "unwatermarked" flow). Before embedding the watermark, the inter-packet delays (IPDs) of the flow, $\tau_i^u = t_{i+1}^u - t_i^u$ are recorded in an IPD database, which is accessible by the watermark detector. The watermark is subsequently embedded by delaying the packets by an amount such that the IPD of the $i$th watermarked packet is $\tau_i^w = \tau_i^u + w_i$. The watermark components $\{w_i\}_{i=1}^n$ take values $\pm a$ with equal probability. The value $a$ is chosen to be small enough so that the artificial jitter caused by watermark embedding is invisible to ordinary users and attackers.[1]

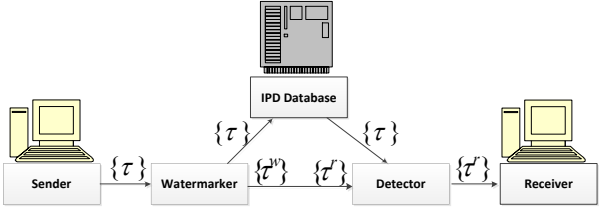

Fig. 1. Model of RAINBOW network flow watermarking system.

In order to apply watermark delays on the flow, output packet $t_i$ is delayed by $w_0 + \sum_{j=1}^{i} w_i$, where $w_0$ is the initial delay applied to the first packet. This results in $\tau_i^w = \tau_i^u + w_i$, as desired. Since we cannot delay a packet for a negative amount of time, $w_0$ must be chosen large enough to prevent this from happening. Since the sequence $w_i$ is generated from a random seed, the watermarker can calculate all of the partial sums $\sum_{j=1}^{i-1} w_i$ in advance and adjust $w_0$ accordingly. If a particular random seed requires a very large initial delay $w_0$, a different seed can be chosen.

As the flow traverses the network, it accumulates extra delays. Let $d_i$ be the delay that the packet accumulates by the time it reaches the watermark detector; i.e., the packet is received at the detector at time $t_i^r = t_i^w + d_i$. The IPD values at the detector are then:

$$\tau_i^r = t_{i+1}^r - t_i^r = \tau_i^u + w_i + \delta_i \qquad (1)$$

where $\delta_i = d_{i+1} - d_i$ is the jitter present in the network.

As mentioned before, the RAINBOW scheme is non-blind and therefore the detector has access to the IPD database where the unwatermarked flows are recorded. Given an observed flow at the detector with IPDs $\tau^r$ and a previously recorded flow $\tau^u$, the detector must decide whether the two flows are linked or not. In the next section we derive the optimum datectors for the RAINBOW watermarks according to the LRT ruls. We also derive the optimum passive detectors, showing that the RAINBOW watermark performs significantly better than passive traffic analysis for correlated network flows.

## 4 DETECTION APPROACHES

RAINBOW is the first *non-blind* flow watermarking scheme. Non-blind watermarking inherits similar scalability issues from the passive traffic analysis. In this section, we show how non-blind watermarking improves

---

the traffic analysis performance as compared to the traditional passive traffic analysis.

We derive *optimum* Likelihood Ratio Test (LRT) detectors for the RAINBOW watermarking scheme for different traffic models, and compare its detection performance with those of optimum passive detectors. We show that RAINBOW outperforms passive traffic analysis for different traffic models; this confirms what we expect intuitively from information theory, as a non-blind watermark detector has access to more information (the watermark and the IPDs), compared to a passive detector which only has access to the IPDs. We also show that the RAINBOW detector is reliable in different models, while the optimum passive detector fails in some scenarios.

As the extreme models, we perform our detection analysis for two traffic models:

- traffic model A: independent flows with i.i.d. inter-packet delays, and,
- traffic model B: completely-correlated flows.

As it is infeasible to evaluate the detection performance for all different traffic models, we discuss the detection performance for these two traffic models, and consider any real-world network flow to lie between these two extreme models. We show that an active detector, i.e., RAINBOW, is reliable for different models, while a passive detector fails for certain traffic models.

### 4.1 Detection primitives

We use *hypothesis testing* [17] to analyze the detection performance of active and passive detectors. For an active detector, we aim to distinguish between the two following hypotheses:

- $H_0$ (*null hypothesis*): the received flow with IPDs $\tau^r$ is a new, unwatermarked flow, unlinked to the flow with IPDs $\tau$,         *and*,
- $H_1$: $\tau^r$ is the result of a flow with original IPDs $\tau$ being watermarked and passed through the network.[1]

Also, for a passive detector we consider the following hypothesis testing problem:

- $H_0$ (*null hypothesis*): the received flow with IPDs $\tau^r$ is a new flow, unlinked to $\tau$ (the IPDs of another received flow),         *and*,
- $H_1$: $\tau^r$ is the result of $\tau$ passing through the network.

We find the *optimum* likelihood-ratio tests (LRT) of these hypothesis testing problems. For any received flow with $\tau^r$ IPDs, an LRT test evaluates a test metric for the IPDs, $T[\tau^r]$, and compares it with a *detection threshold* $\eta$; if $T[\tau^r] \geq \eta$, the received flow is said to be *linked* to the one in the detector's database (with IPDs of $\tau$). We

---

1. Throughout this paper, by attacker we mean the attacker to the watermarking scheme.

1. Note that there is another possibility, namely that $\tau^r$ is a *watermarked* flow, but not corresponding to $\tau$. However, we ignore this case because errors in this scenario do not matter: if the flow is said to be watermarked, then the detection algorithm is correct, and if it is said to be unwatermarked, it will later be tested against the correct $\tau$.

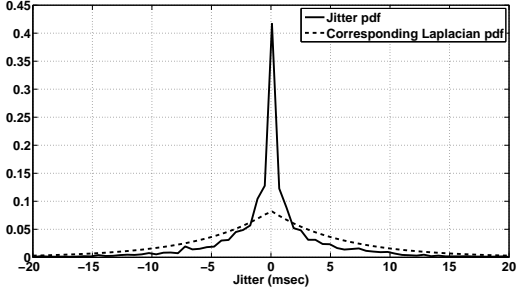

Fig. 2. A comparison of observed jitter and a fitted Laplace distribution.

can therefore express the false positive and false negative rates of the detector as:

$$P_{\text{FP}} = P(T[\boldsymbol{\tau}^r | H_0] \geq \eta) \tag{2}$$

$$P_{\text{FN}} = P(T[\boldsymbol{\tau}^r | H_1] < \eta) \tag{3}$$

### 4.2 Network jitter model

We will model network delays as i.i.d. exponential, which implies that the jitter (difference of two delays) is i.i.d. according to a zero-mean Laplace distribution denoted by $Lap(0, b_\delta)$, where $2b_\delta^2$ is the variance of the jitter. Of course, in a real network, delays will have some correlation; we compare the probability density function (PDF) of real observed jitter on a connection over Planet-Lab [16] with a best-fit Laplace distribution in Figure 2. We can see that the real PDF has greater support at 0, and the Laplace distribution has a heavier tail. This means that our analysis of error rates will be conservative, since 0 jitter will result in no error for our detection scheme. We have also conducted similar experiments with the same results on Tor anonymous network [23] to consider the other application of watermarking.

### 4.3 Traffic model A: independent flows, i.i.d. IPDs

In this model, we assume that the candidate flows are independent. Also, each flow has i.i.d. IPDs, i.e., the flow is modeled with a Poisson process. This represents a good model for non-interactive network flows.

#### 4.3.1 Passive detection (PASSV scheme)

In this section, we find the optimum likelihood ratio (LRT) passive detector for the traffic model A. Suppose that the flow with IPDs $\boldsymbol{\tau}$ is known to the detector. Detector will need to check if it is correlated with some received flow $\boldsymbol{\tau}^*$, where $\boldsymbol{\tau}$ and $\boldsymbol{\tau}^*$ are independent. So, in this case the hypothesis testing problem is:

$$\begin{cases} H_0: & \tau_i^r = \tau_i^* + \delta_i^0 \\ H_1: & \tau_i^r = \tau_i + \delta_i^1 \end{cases} \tag{4}$$

where $\delta^0$ and $\delta^1$ represent the network jitter. Based on our measurements over the Planetlab we model the network jitter with an i.i.d. Laplacian distribution $Lap(0, b)$ (see Section 4.2).

In order to find the optimum LRT detector, we first need to find the PDF function of $\tau_i^r$ in different hypotheses, i.e., $p_i(\cdot)$ for hypothesis $H_i$. As the model A suggests, we model the IPDs $\boldsymbol{\tau}^*$ as i.i.d. exponential distribution. So, in hypothesis $H_0$ the received signal $\tau_i^r$ is the summation of a Laplacian and an exponential random variable; we use Lemma 3 in Appendix A to find $p_0(\cdot)$:

$$p_0(\tau_i^r) = \begin{cases} \frac{\lambda}{2(\lambda b - 1)} e^{-\frac{\tau_i^r}{b}} + \frac{\lambda}{1 - \lambda^2 b^2} e^{-\lambda \tau_i^r} & y_i \geq 0 \\ \frac{\lambda}{2(\lambda b + 1)} e^{\frac{\tau_i^r}{b}} & y_i < 0 \end{cases} \tag{5}$$

In the case of $H_1$, since the $\tau_i$ is known to the detector, we can model $\tau_i^r$ as a Laplacian distribution with mean $\tau_i$. So:

$$p_1(\tau_i^r) = \frac{1}{2b} e^{-\frac{|\tau_i^r - \tau_i|}{b}} \tag{6}$$

Note that even though the real-world IPDs can never be negative, the densities $p_0$ and $p_1$ return a non-zero density for negative values of the IPDs. In fact, this is due to the approximation we make in modeling the network jitter as a two-sided Laplacian distribution, and its effect is very small for ordinary network flows based on our simulations [15].

Having the densities $p_0$ and $p_1$, we derive the optimum detector based on the likelihood ratio test to be:

$$L(\boldsymbol{\tau}^r) \underset{H_0}{\overset{H_1}{\gtrless}} e^\eta \tag{7}$$

where $\eta$ is the LRT detection threshold and

$$L(\boldsymbol{\tau}^r) = \prod L_i(\tau_i^r) \tag{8}$$

$$L_i(\tau_i^r) = \frac{p_1(\tau_i^r)}{p_0(\tau_i^r)} \tag{9}$$

We define $\eta_n = \eta/n$ as the *normalized detection threshold*. A value of of $\eta_n = 0$ results in a MiniMax detector.

4.3.1.1 Detection performance: Let us consider the case where the detector uses the PASSV detection scheme in order to link a received flow with IPDs $\boldsymbol{\tau}^r$ to a known flow with IPDs $\boldsymbol{\tau}$, i.e., a registered flow. Considering the assumptions made in the traffic model A, i.e., the IPDs being i.i.d., we use Lemma 1 (part b) in Appendix A to find the false positive ($P_{FP}$) and false negative ($P_{FN}$) error rates of the PASSV detector:

$$P_{FP}^{\boldsymbol{\tau}} \leq \prod_{i=1}^{n} e^{-(s\eta_n - \mu_{0,i}^{\tau_i}(s))} \tag{10}$$

$$P_{FN}^{\boldsymbol{\tau}} \leq \prod_{i=1}^{n} e^{-((s-1)\eta_n - \mu_{0,i}^{\tau_i}(s))} \tag{11}$$

where $0 < s < 1$ and:

$$\mu_{0,i}^{\tau_i}(s) = \ln \int p_0^{1-s}(\tau_i^r) p_1^s(\tau_i^r) d\tau_i^r \tag{12}$$

The error probabilities of $P_{FN}^{\boldsymbol{\tau}}$ and $P_{FP}^{\boldsymbol{\tau}}$ correspond to a fixed known IPDs sequence, $\boldsymbol{\tau}$. The overall false errors are evaluated by averaging $P_{FP}^{\boldsymbol{\tau}}$ and $P_{FN}^{\boldsymbol{\tau}}$ with respect to $\boldsymbol{\tau}$:

$$P_{FP} = E_{\boldsymbol{\tau}}\{P_{FP}^{\boldsymbol{\tau}}\} \tag{13}$$

$$\leq \prod_{i=1}^{n} E_{\tau_i}\left\{e^{-(s\eta_n - \mu_{0,i}^{\tau_i}(s))}\right\} \tag{14}$$

$$= \left(\int_0^\infty e^{-(s\eta_n - \mu_{0,1}^{\tau_1}(s))}\lambda e^{-\lambda\tau_1}d\tau_1\right)^n \tag{15}$$

$$P_{FN} = E_{\boldsymbol{\tau}}\{P_{FN}^{\boldsymbol{\tau}}\} \tag{16}$$

$$\leq \prod_{i=1}^{n} E_{\tau_i}\left\{e^{-((s-1)\eta_n - \mu_{0,i}^{\tau_i}(s))}\right\} \tag{17}$$

$$= \left(\int_0^\infty e^{-((s-1)\eta_n - \mu_{0,1}^{\tau_1}(s))}\lambda e^{-\lambda\tau_1}d\tau_1\right)^n \tag{18}$$

We can represent the upper bounds of these false errors as:

$$P_{FP} \leq e^{-n\cdot E_{FP}(s,\eta_n)} \tag{19}$$
$$P_{FN} \leq e^{-n\cdot E_{FN}(s,\eta_n)} \tag{20}$$

where

$$E_{FP}(s,\eta_n) = -\ln\left(\int_0^\infty e^{-(s\eta_n - \mu_{0,1}^{\tau_1}(s))}\lambda e^{-\lambda\tau_1}d\tau_1\right) \tag{21}$$

$$E_{FN}(s,\eta_n) = -\ln\left(\int_0^\infty e^{-((s-1)\eta_n - \mu_{0,1}^{\tau_1}(s))}\lambda e^{-\lambda\tau_1}d\tau_1\right) \tag{22}$$

$$(0 < s < 1)$$

For each detection threshold $\eta_n$, we find the tightest exponent bounds $E_{FP}^*(\eta_n)$ and $E_{FN}^*(\eta_n)$ such that:

$$E_{FP}^*(\eta_n) = \max_{0<s<1} E_{FP}(s,\eta_n) \tag{23}$$

$$E_{FN}^*(\eta_n) = \max_{0<s<1} E_{FN}(s,\eta_n) \tag{24}$$

### 4.3.1.2 Analysis results:
We use Mathematica 7.0 to evaluate the false error exponents of (23) and (24). The parameters used for the simulations are $b = 10^{-2}sec$ and $\lambda = 5pps$, borrowed from [15]. Figure 3 plots the tightest bounds for the error exponents of $E_{FP}^*(\eta_n)$ and $E_{FN}^*(\eta_n)$ for different thresholds of $\eta_n$. Note that the optimum $s$ varies with the decision threshold. For $\eta_n = 0$ the false positive and false negative errors are equal; we name this error rate as the *Cross-Over Error Rate* (COER). For the mentioned setting of the variables the COER exponent of the PASSV detector is equal to $1.06396$.

### 4.3.2 Active detection (ACTV scheme)
In this section, we find the optimum LRT detector for the RAINBOW non-blind watermark for the traffic model A. We have the following hypothesis testing problem:

$$\begin{cases} H_0 : \tau_i^r = \tau_i^* + \delta_i \\ H_1 : \tau_i^r = \tau_i + w_i + \delta_i \end{cases} \tag{25}$$

where $\tau_i$'s are the IPDs registered in the IPD database, and $\tau_i^*$'s are the IPDs of an independent flow. As before, in order to find the optimum LRT detector we need to find the distribution of $\tau_i^r$ in different hypotheses. Using Lemma 3 in Appendix B we find the corresponding PDF function under $H_0$ as:

$$p_0(\tau_i^r) = \begin{cases} \frac{\lambda}{2(\lambda b - 1)}e^{-\frac{\tau_i^r}{b}} + \frac{\lambda}{1-\lambda^2 b^2}e^{-\lambda\tau_i^r} & \tau_i^r \geq 0 \\ \frac{\lambda}{2(\lambda b+1)}e^{\frac{\tau_i^r}{b}} & \tau_i^r < 0 \end{cases} \tag{26}$$

Since $\tau_i$ and $w_i$ are known to the detector, we find the PDF in hypothesis $H_1$ as the following:

$$p_1(\tau_i^r) = \frac{1}{2b}e^{-\frac{|\tau_i^r - \tau_i - w_i|}{b}} \tag{27}$$

So, the optimum detector based on the likelihood ratio test is:

$$L(\boldsymbol{\tau}^r) \gtrless_{H_0}^{H_1} e^\eta \tag{28}$$

where $\eta$ is the LRT detection threshold and

$$L(\boldsymbol{\tau}^r) = \prod L_i(\tau_i^r) \tag{29}$$

$$L_i(\tau_i^r) = \frac{p_1(\tau_i^r)}{p_0(\tau_i^r)} \tag{30}$$

### 4.3.2.1 Detection performance:
As before, considering the independence of the IPDs and also the watermark bits we use Lemma 1 (part b) in Appendix A to find the error probabilities of the ACTV detector for a given $\boldsymbol{\tau}$ and $\mathbf{w}$:

$$P_{FP}^{\boldsymbol{\tau},\mathbf{w}} \leq \prod_{i=1}^{n} e^{-(s\eta_n - \mu_{0,i}^{\tau_i,w_i}(s))} \tag{31}$$

$$P_{FN}^{\boldsymbol{\tau},\mathbf{w}} \leq \prod_{i=1}^{n} e^{-((s-1)\eta_n - \mu_{0,i}^{\tau_i,w_i}(s))} \tag{32}$$

where $0 < s < 1$, and:

$$\mu_{0,i}^{\tau_i,w_i}(s) = \ln\int p_0^{1-s}(\tau_i^r)p_1^s(\tau_i^r)d\tau_i^r \tag{33}$$

As $P_{FN}^{\boldsymbol{\tau},\mathbf{w}}$ and $P_{FP}^{\boldsymbol{\tau},\mathbf{w}}$ correspond to a fixed IPDs sequence $\boldsymbol{\tau}$ and the watermark $\mathbf{w}$, we evaluate the overall false errors by averaging $P_{FP}^{\boldsymbol{\tau},\mathbf{w}}$ and $P_{FN}^{\boldsymbol{\tau},\mathbf{w}}$ with respect to $\boldsymbol{\tau}$ and $\mathbf{w}$:

$$P_{FP} = E_{\mathbf{w}}E_{\boldsymbol{\tau}}\{P_{FP}^{\boldsymbol{\tau},\mathbf{w}}\} \tag{34}$$

$$\leq \prod_{i=1}^{n} E_{w_i}E_{\tau_i}\left\{e^{-(s\eta_n - \mu_{0,i}^{\tau_i,\mathbf{w}}(s))}\right\} \tag{35}$$

$$= \left(\frac{1}{2}\sum_{w_1=0}^{1}\int_0^\infty e^{-(s\eta_n - \mu_{0,1}^{\tau_1,w_1}(s))}\lambda e^{-\lambda\tau_1}d\tau_1\right)^n \tag{36}$$

$$P_{FN} = E_{\mathbf{w}}E_{\boldsymbol{\tau}}\{P_{FN}^{\boldsymbol{\tau},\mathbf{w}}\} \tag{37}$$

$$\leq \prod_{i=1}^{n} E_{w_i}E_{\tau_i}\left\{e^{-((s-1)\eta_n - \mu_{0,i}^{\tau,\mathbf{w}}(s))}\right\} \tag{38}$$

$$= \left(\frac{1}{2}\sum_{w_1=0}^{1}\int_0^\infty e^{-((s-1)\eta_n - \mu_{0,1}^{\tau_1,w_1}(s))}\lambda e^{-\lambda\tau_1}d\tau_1\right)^n \tag{39}$$

The approximated upperbounds can be formulated as:

$$P_{FP} \leq e^{-n \cdot E_{FP}(s, \eta_n)} \tag{40}$$

$$P_{FN} \leq e^{-n \cdot E_{FN}(s, \eta_n)} \tag{41}$$

where

$$E_{FP}(s, \eta_n) = -\ln\left(\frac{1}{2} \sum_{w_1=0}^{1}\right.$$
$$\left. \int_0^\infty e^{-(s\eta_n - \mu_{0,1}^{\tau_1, w_1}(s))} \lambda e^{-\lambda \tau_1} d\tau_1\right) \tag{42}$$

$$E_{FN}(s, \eta_n) = -\ln\left(\frac{1}{2} \sum_{w_1=0}^{1}\right.$$
$$\left. \int_0^\infty e^{-((s-1)\eta_n - \mu_{0,1}^{\tau_1, w_1}(s))} \lambda e^{-\lambda \tau_1} d\tau_1\right) \tag{43}$$
$$(0 < s < 1)$$

Finally, the tightest bounds for each $\eta_n$ are found by maximizing the error exponents with respect to the parameter $s$:

$$E_{FP}^*(\eta_n) = \max_{0 < s < 1} E_{FP}(s, \eta_n) \tag{44}$$

$$E_{FN}^*(\eta_n) = \max_{0 < s < 1} E_{FN}(s, \eta_n) \tag{45}$$

4.3.2.2 Analysis results: Using Mathematica 7.0 we evaluate the false error exponents of (44) and (45). As before, we use the parameters $b = 10^{-2}sec$, $a = 10^{-2}sec$, and $\lambda = 5pps$ for the simulations. Figure 4 plots the tightest bounds for the error exponents of $E_{FP}^*(\eta_n)$ and $E_{FN}^*(\eta_n)$ for different thresholds of $\eta_n$. The COER exponent occurs for $\eta_n = 0$ and is equal to 1.06828, which is slightly better compared to that of the PASSV detector evaluated before, i.e., 1.06396.

## 4.4 Traffic model B: correlated flows, correlated IPDs

As the other extreme of traffic models we investigate the traffic model with correlated IPDs. We consider the case where all of the network flows have the same IPDs, e.g., for any two flows with IPDs $\boldsymbol{\tau}^*$ and $\boldsymbol{\tau}$ we have that $\tau_i^* = \tau_i = C_i$ for all $i$. In particular, this model captures the behavior of a number of widely used traffic types, including file transfers, browsing the same websites, etc.

### 4.4.1 Passive detection

In this model, a passive detection faces the following hypothesis testing problem:

$$\begin{cases} H_0 : & \tau_i^r = \tau_i^* + \delta_i \\ H_1 : & \tau_i^r = \tau_i + \delta_i \end{cases} \tag{46}$$

where $\tau_i^* = \tau_i = C_i$. The optimum LRT detector for this problem is the random guessing:

$$L(\boldsymbol{\tau}^r) = RND \tag{47}$$

where $RND$ is a uniform random variable. The detection rule is:

$$L(\boldsymbol{\tau}^r) \gtrless_{H_0}^{H_1} e^\eta \tag{48}$$

4.4.1.1 Detection performance: Since the detector is based on random guessing, the false errors are as followed:

$$P_{FP} = p \tag{49}$$

$$P_{FN} = 1 - p \tag{50}$$

where $0 \leq p \leq 1$ is determined by the choice of $\eta$.

### 4.4.2 Active detection (SLCorr scheme)

In this case, we have the following hypothesis testing problem:

$$\begin{cases} H_0 : & \tau_i^r = \tau_i^* + \delta_i \\ H_1 : & \tau_i^r = \tau_i + w_i + \delta_i \end{cases} \tag{51}$$

Since $\tau_i^* = \tau_i = C_i$, this can be reduced to the following hypothesis testing:

$$\begin{cases} H_0 : & y_i = \delta_i \\ H_1 : & y_i = w_i + \delta_i \end{cases} \tag{52}$$

where $y_i = \tau_i^r - \tau_i$. The optimum LRT detector for this problem can be found considering the distribution of $y_i$ in different hypotheses:

$$p_0^i(y_i) = \frac{1}{2b} e^{-\frac{|y_i|}{b}} \tag{53}$$

$$p_1^i(y_i) = \frac{1}{2b} e^{-\frac{|y_i - w_i|}{b}} \tag{54}$$

So, we can derive the LRT detection metric as:

$$L_i(y_i) = \frac{p_1^i(y_i)}{p_0^i(y_i)} \tag{55}$$

which can be expressed as:

$$\ln L_i(y_i) = \frac{1}{b}(|y_i| - |y_i - w_i|) \tag{56}$$

$$= \frac{2}{b} f^{SL}\left(y_i - \frac{w_i}{2}\right).sgn(w_i) \tag{57}$$

$f^{SL}(\cdot)$ is a *soft-limiter* with breakpoints at $-\frac{a}{2}$ and $+\frac{a}{2}$ ($a$ is the watermark amplitude as defined before):

$$f^{SL}(x) = \begin{cases} +\frac{a}{2} & x \geq +\frac{a}{2} \\ x & -\frac{a}{2} < x < +\frac{a}{2} \\ -\frac{a}{2} & x \leq -\frac{a}{2} \end{cases} \tag{58}$$

We can reformulate the optimum detection rule as:

$$D(\mathbf{y}) \gtrless_{H_0}^{H_1} \eta \tag{59}$$

where

$$D(\mathbf{y}) = \sum_{i=1}^n D_i(y_i) \tag{60}$$

and

$$D_i(y_i) = \frac{b}{2} \ln L_i(y_i)$$
$$= f^{SL}\left(y_i - \frac{w_i}{2}\right).sgn(w_i) \tag{61}$$

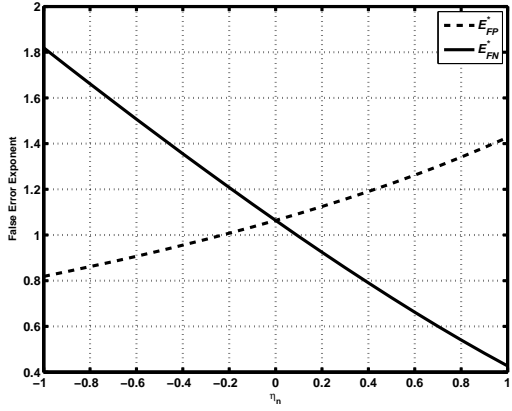

Fig. 3. Analytical error exponents $E_{FP}^*(\eta_n)$ and $E_{FN}^*(\eta_n)$ of the PASSV detection scheme for different values of $\eta_n$ (traffic model A). ($b = 10^{-2}sec$, $\lambda = 5pps$)

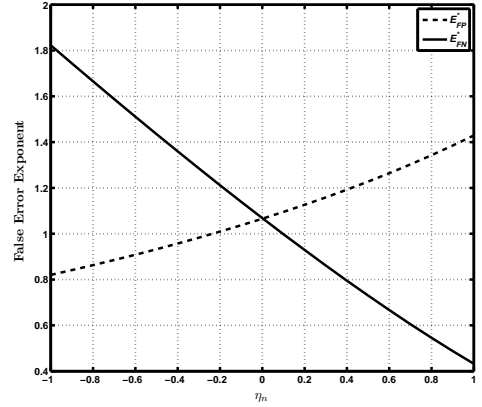

Fig. 4. Analytical error exponents $E_{FP}^*(\eta_n)$ and $E_{FN}^*(\eta_n)$ of the ACTV detection scheme for different values of $\eta_n$ (traffic model A). ($b = 10^{-2}sec$, $\lambda = 5pps$)

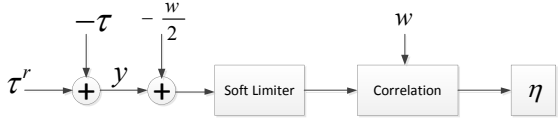

Fig. 5. Block diagram of the SLCorr detection scheme.

We call this detector *SLCorr*, as it is composed of a soft limiter followed by a correlation block. From a communications point of view, the soft-limiter is useful in reducing the signal detection noise in channels with a Laplacian distributed noise. We will use this as the detection scheme for the RAINBOW watermark, as would be discussed later. Figure 5 shows the block diagram of the *SLCorr* detector. *SLCorr* is a MiniMax detector for a detection threshold of $\eta = 0$.

4.4.2.1 Detection performance: The SLCorr test metric is given in (59) to (61). Let us define $f_0^i(\cdot)$ and $f_1^i(\cdot)$ as the PDF of $x_i = y_i - \frac{w_i}{2}$ in hypothesis $H_0$ and $H_1$, respectively. We have that:

$$f_0^i(x_i) = \frac{1}{2b} e^{-\frac{|x_i + \frac{w_i}{2}|}{b}} \qquad (62)$$

$$f_1^i(x_i) = \frac{1}{2b} e^{-\frac{|x_i - \frac{w_i}{2}|}{b}} \qquad (63)$$

Based on these, we can evaluate $p_0(\cdot)$ and $p_1(\cdot)$, namely the PDF of $D_i(y_i)$ under hypothesis $H_0$ and $H_1$, respectively:

$$p_0(D_i) = \begin{cases} \frac{1}{2} e^{-\frac{a}{b}} & D_i = +\frac{a}{2} \\ \frac{1}{2b} e^{-\frac{D_i + \frac{a}{2}}{b}} & -\frac{a}{2} < D_i < \frac{a}{2} \\ \frac{1}{2} & D_i = -\frac{a}{2} \end{cases} \qquad (64)$$

$$p_1(D_i) = \begin{cases} \frac{1}{2} & D_i = +\frac{a}{2} \\ \frac{1}{2b} e^{\frac{D_i - \frac{a}{2}}{b}} & -\frac{a}{2} < D_i < \frac{a}{2} \\ \frac{1}{2} e^{-\frac{a}{2}} & D_i = -\frac{a}{2} \end{cases} \qquad (65)$$

Considering that the distributions $p_0(D_i)$ and $p_1(D_i)$ are i.i.d. with $i$ we use the Chernof bound (part (c) of Lemma 1 in Appendix A) to find the error probabilities of the SLCorr detector:

$$P_{FP} \le e^{-n(s\eta_n - \mu_0(s))} \qquad (\forall s > 0) \qquad (66)$$

$$\mu_0(s) = \mu_{D_i|H_0}(s)$$

$$P_{FN} \le e^{-n(s\eta_n - \mu_1(s))} \qquad (\forall s < 0) \qquad (67)$$

$$\mu_1(s) = \mu_{D_i|H_1}(s)$$

where $\eta_n = \eta/n$ is the normalized detection threshold. We have that:

$$\mu_0(s) = \mu_{D_i|H_0}(s) = \ln \int_{-\infty}^{\infty} e^{sx} p_0(x) dx$$
$$= \ln \left[ \frac{sb}{2(sb-1)} e^{-\frac{a}{b}} e^{s\frac{a}{2}} + \frac{sb-2}{2(sb-1)} e^{-s\frac{a}{2}} \right] \qquad (68)$$

and,

$$\mu_1(s) = \mu_{D_i|H_1}(s) = \ln \int_{-\infty}^{\infty} e^{sx} p_1(x) dx$$
$$= \ln \left[ \frac{sb}{2(sb+1)} e^{-\frac{a}{b}} e^{-s\frac{a}{2}} + \frac{sb+2}{2(sb+1)} e^{s\frac{a}{2}} \right] \qquad (69)$$

We can express the above $P_{FP}$ and $P_{FN}$ false errors as:

$$P_{FP} \le e^{-n \cdot E_{FP}(s,\eta_n)} \qquad (70)$$

$$P_{FN} \le e^{-n \cdot E_{FN}(s,\eta_n)} \qquad (71)$$

where

$$E_{FP}(s,\eta_n) = s\eta_n - \mu_0(s) \qquad (s > 0) \qquad (72)$$

$$E_{FN}(s,\eta_n) = s\eta_n - \mu_0(s) \qquad (s < 0) \qquad (73)$$

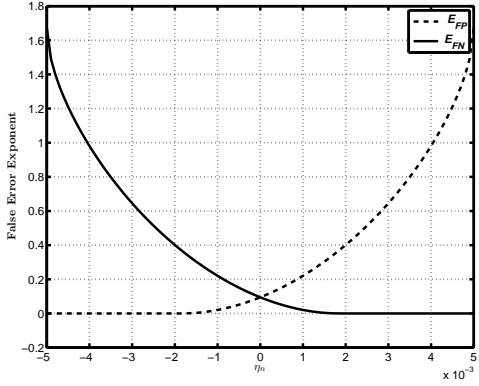

Fig. 6. Analytical error exponents $E_{FP}^*(\eta_n)$ and $E_{FN}^*(\eta_n)$ of SLCorr for different values of $\eta_n$ (traffic model B). ($b = 10^{-2}sec$, $a = 10^{-2}sec$)

Finally, the tightest bounds for each $\eta_n$ are found by maximizing error exponents with respect to the $s$ parameter:

$$E_{FP}^*(\eta_n) = \max_{s>0} E_{FP}(s, \eta_n) \qquad (74)$$

$$E_{FN}^*(\eta_n) = \max_{s<0} E_{FN}(s, \eta_n) \qquad (75)$$

4.4.2.2  Analysis results: We use Mathematica 7.0 to evaluate the false error exponents of (74) and (75). The parameters used for the simulations are $b = 10^{-2}sec$ and $a = 10^{-2}sec$. Figure 6 plots the tightest bounds for the error exponents of $E_{FP}^*(\eta_n)$ and $E_{FN}^*(\eta_n)$ for different thresholds of $\eta_n$. The COER exponent occurs for $\eta_n = 0$ and is equal to 0.0945.

### 4.5  Discussion

Above, we derived the optimum passive and active detectors for the traffic analysis problem and evaluated their performance by finding the Chernoff upperbounds of their false error rates. In this section, we use the *asymptotic relative efficiency* (ARE) as a tool to compare their detection performances.

The asymptotic relative efficiency (ARE) is a measure for comparing two discrete-time detection schemes. For two discrete detection schemes $S_1$ and $S_2$ the ARE metric is defined as $ARE_{S_1,S_2} = \lim_{n\to\infty} n_2/n$, where $n$ is the number of $S_1$'s samples. The $n_2$ parameter is the smallest number of $S_2$ samples that results in $S_2$'s error rate to be smaller than or equal to the error rate of $S_1$ (with $n$ samples). An ARE metric of $ARE_{S_1,S_2} > 1$ depicts that $S_1$ is asymptotically more efficient than $S_2$. Chernoff [24] finds the ARE metric of two detectors $S_1$ and $S_2$ using their Chernoff error upperbounds as:

$$ARE_{S_1,S_2} = E_1/E_2 \qquad (76)$$

where $E_1$ and $E_2$ are the error exponents of the Chernoff upperbounds for $S_1$ and $S_2$ detectors, respectively.

Using the analysis results from Sections 4.3 and 4.4 we can derive the ARE metric of the optimum passive and active detectors for the two traffic models as:

$$ARE_{PASSV,ACTV}|A = 1.06396/1.06828 \approx 0.996 \qquad (77)$$

$$ARE_{RND,SLCorr}|B = 0/0.0945 = 0 \qquad (78)$$

This asserts that the optimum active detector outperforms the optimum passive detector in both traffic models A and B (which is intuitively expected from information theory). As an important observation, we see that the active detector's advantage is very small for the traffic model A, however, the active detector significantly outperforms the optimum passive detector in traffic model B, i.e., the correlated traffic. In other words, the active detector provides very good detection performance for different traffic models, however, the passive detection is very poor for the more correlated network traffic. Later in this section, we sh

In the rest of this section we analyze the performance of the SLCorr scheme under the traffic model A, showing that even though SLCorr is not the optimum detector for the traffic model A, however, it provides very good detection performance under this model. Based on this, we choose SLCorr as the sole detector for RAINBOW, regardless of the behavior of the network flows. This simplifies the watermark detection, as real-world traffic are combinations of the models A and B, and the detection can be performed regardless of the type of the received traffic. We also analyze the performance of PASSV and ACTV detectors under traffic model B, showing their inefficiency in this model.

### 4.5.1  SLCorr Detection performance for traffic model A

The SLCorr scheme is the optimum active detector for traffic model B, but not the traffic model A. In this section we show that SLCorr achieves a good detection performance even under traffic model A, allowing a system designer to use it as the sole detection scheme regardless of the type of the traffic. SLCorr faces the following hypothesis testing under the traffic model A:

$$\begin{cases} H_0 : \tau_i^r = \tau_i^* + \delta_i \\ H_1 : \tau_i^r = \tau_i + w_i + \delta_i \end{cases} \qquad (79)$$

Considering SLCorr's detection metric, given in (59) to (61), one can rewrite the hypothesis testing problem as:

$$\begin{cases} H_0 : y_i = \tau_i^* + \delta_i - \tau_i \\ H_1 : y_i = w_i + \delta_i \end{cases} \qquad (80)$$

where $y_i = \tau_i^r - \tau_i$. Let us assume $f_i^0(\cdot)$ and $f_i^1(\cdot)$ as the PDF functions of $y_i|H_0$ and $y_i|H_1$, respectively. We have that:

$$y_i|H_1 \sim Lap(w_i, b) \qquad (81)$$

$$f_i^1(y_i) = \frac{1}{2b}e^{-\frac{|y_i - w_i|}{b}} \qquad (82)$$

Also using Lemma2 in Appendix B:

$$\delta_i \sim Lap(0, b) \tag{83}$$

$$(\tau_i^* - \tau_i) \sim Lap(0, 1/\lambda) \tag{84}$$

$$f_i^0(y_i) = \frac{b\lambda}{2(1 - b^2\lambda^2)} \left( \frac{1}{b} e^{-\lambda|y_i|} - \lambda e^{-\frac{1}{b}|y_i|} \right) \tag{85}$$

Now, let us define $p_0(\cdot)$ and $p_1(\cdot)$ as the PDF functions of $D_i(y_i)$ under hypotheses $H_0$ and $H_1$, respectively. We derive $p(\cdot)$ as:

$$p_0(D_i) = \begin{cases} \frac{b^2\lambda^2}{2(1-b^2\lambda^2)} \left( \frac{1}{b^2\lambda^2} e^{-\lambda a} - e^{-\frac{a}{b}} \right) & D_i = +\frac{a}{2} \\ \frac{b\lambda}{2(1-b^2\lambda^2)} \left( \frac{1}{b} e^{-\lambda(D_i + a/2)} \right. \\ \quad \left. -\lambda e^{-\frac{1}{b}(D_i + a/2)} \right) & -\frac{a}{2} < D_i < \frac{a}{2} \\ \frac{1}{2} & D_i = -\frac{a}{2} \end{cases} \tag{86}$$

Also, using (82) we derive $p_1(\cdot)$ as:

$$p_1(D_i) = \begin{cases} \frac{1}{2} & D_i = +\frac{a}{2} \\ \frac{1}{2b} e^{\frac{D_i - \frac{a}{2}}{b}} & -\frac{a}{2} < D_i < \frac{a}{2} \\ \frac{1}{2} e^{-\frac{a}{b}} & D_i = -\frac{a}{2} \end{cases} \tag{87}$$

Based on the $p_0(\cdot)$ and $p_1(\cdot)$ distributions and using the Chernoff bounds for signal detection (part c of Lemma 1 in Appendix A) we find the error probabilities of the detector to be:

$$P_{FP} \leq e^{-n(s\eta_n - \mu_0(s))} \qquad (\forall s > 0) \tag{88}$$

$$\mu_0(s) = \mu_{D_i|H_0}(s)$$

$$P_{FN} \leq e^{-n(s\eta_n - \mu_1(s))} \qquad (\forall s < 0) \tag{89}$$

$$\mu_1(s) = \mu_{D_i|H_1}(s)$$

where we have:

$$\mu_0(s) = \mu_{D_i|H_0}(s) = \ln \int_{-\infty}^{\infty} e^{sx} p_0(x) dx \tag{90}$$

$$= \ln \left[ \frac{b^2\lambda^2}{2(1 - b^2\lambda^2)} \left[ \frac{s}{b^2\lambda^2(s - \lambda)} e^{sa/2} e^{-\lambda a} \right. \right.$$
$$+ \frac{sb}{1 - sb} e^{sa/2} e^{-a/b}$$
$$\left. \left. + \frac{-2\lambda bs + 2\lambda + sb^2\lambda^2 - b^2\lambda^3 + s^2 b - s}{(s - \lambda)(sb - 1)b^2\lambda^2} e^{-sa/2} \right] \right] \tag{91}$$

and,

$$\mu_1(s) = \mu_{D_i|H_1}(s) = \ln \int_{-\infty}^{\infty} e^{sx} p_1(x) dx \tag{92}$$

$$= \ln \left[ \frac{sb}{2(sb+1)} e^{-\frac{a}{b} - s\frac{a}{2}} + \frac{sb + 2}{2(sb+1)} e^{s\frac{a}{2}} \right] \tag{93}$$

As before, we can express the above $P_{FP}$ and $P_{FN}$ false errors as:

$$P_{FP} \leq e^{-n \cdot E_{FP}(s, \eta_n)} \tag{94}$$

$$P_{FN} \leq e^{-n \cdot E_{FN}(s, \eta_n)} \tag{95}$$

where

$$E_{FP}(s, \eta_n) = s\eta_n - \mu_0(s) \qquad (s > 0) \tag{96}$$

$$E_{FN}(s, \eta_n) = s\eta_n - \mu_0(s) \qquad (s < 0) \tag{97}$$

Finally, the tightest bounds for each $\eta_n$ are found by maximizing the error exponents with respect to the parameter $s$:

$$E_{FP}^*(\eta_n) = \max_{s > 0} E_{FP}(s, \eta_n) \tag{98}$$

$$E_{FN}^*(\eta_n) = \max_{s < 0} E_{FN}(s, \eta_n) \tag{99}$$

4.5.1.1 Analysis results: We use Mathematica 7.0 to evaluate the false error exponents of (98) and (99). The parameters used for the simulations are $b = 10^{-2} sec$, $\lambda = 5pps$ and $a = 10^{-2} sec$. Figure 7 plots the tightest bounds for the error exponents of $E_{FP}^*(\eta_n)$ and $E_{FN}^*(\eta_n)$ for different thresholds of $\eta_n$. The COER exponent occurs for $\eta_n = 9.6 \times 10^4 s$ which is equal to 0.0228. Also, Figure 8 shows the COER exponent with respect to different values of the watermark amplitude, $a$. As we can see, increasing the watermark amplitude improves the detection performance (but reduces the watermark invisibility as discussed in [15]).

### 4.5.2 Detection performance of PASSV and ACTV schemes for traffic model B

As derived before, the PASSV and ACTV schemes are the optimum passive and active detectors for the traffic model A. We show that PASSV and ACTV perform very poor under the traffic model B, i.e., the correlated traffic. This is unlike the SLCorr detector that works good for both of the traffic models.

Under the traffic model B, the PASSV detector faces the hypothesis testing problem of (46) with $\tau_i^* = \tau_i = C_i$. One can see that in this case the PASSV detection rule described in Section 4.3.1 is exactly the same for both $H0$ and $H1$ hypotheses. This means that the false positive error rate of PASSV scheme for correlated flows is equal to its true positive rate, which makes the PASSV scheme equivalent to a random guessing detector. Similarly, for the traffic model B the ACTV scheme deals with the hypothesis testing problem of (51) with $\tau_i^* = \tau_i = C_i$. Our analysis and simulations on Mathematica confirms that the ACTV detection metric results in very close values for the two hypothesis of $H0$ and $H1$, rendering the ACTV detection scheme ineffective for network flows in traffic model B (we skip the details due to the space constraints).

## 5 SIMULATION RESULTS

In this section, we evaluate the performance of the three detection schemes introduced before, i.e., SLCorr, ACTV, and PASSV, through simulating them over real-world traffic. We show that SLCorr outperforms the other detectors dealing with real-world network flows, due to the intrinsic correlations among the real-world network flows. We use the CAIDA network traces gathered January 2009 [25] for our simulations. For our simulations, we have implemented the detection schemes in C++. From the CAIDA traces we extract three types of network flows for our simulations: TCP ports of 443 (HTTPS), 25 (SMTP), and 22 (SSH). We only select flows with rates lower than 30pps (this is because the parameters of the optimum detectors depend on the rate of the flows). In all of the simulations, the detectors use the detection thresholds derived through analysis in the previous sections, i.e., 0.001 for SLCorr, 0 for ACTV, and 0 for PASSV.

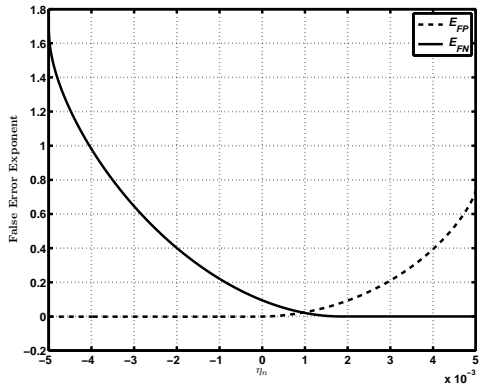

Fig. 7. Analytical error exponents $E^*_{FP}(\eta_n)$ and $E^*_{FN}(\eta_n)$ of SLCorr for different values of $\eta_n$ (traffic model A). ($b = 10^{-2}sec$, $\lambda = 5pps$, $a = 10^{-2}sec$)

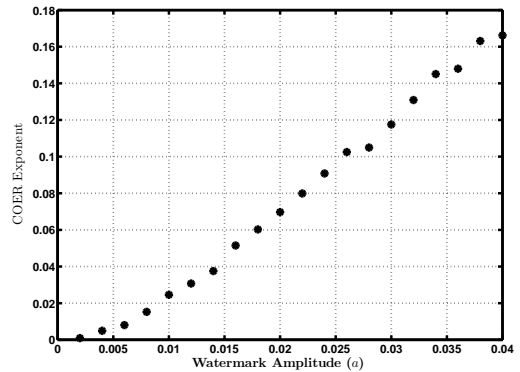

Fig. 8. The COER error exponent of SLCorr in traffic model A for different watermark amplitudes.

In the first set of our simulations, we evaluate the false positive error rate of the three detection schemes for network flows mentioned above. For each detection scheme, we run the detection algorithm for 10000 different pairs of network flows. In order to show the effect of number of packets in the detection performance, we run the experiments for four different values of the $N$ parameter, i.e., 25, 50, 100, and 200. Tables 1, 2, and 3 show the false positive rates of the experiments along with some statistics on the detection metrics for three TCP ports of 443, 25, and 22, respectively. Results show that in most of the cases the SLCorr scheme results in smaller false positive errors compared to the ACTV and PASSV schemes. This is because the real network flows are deviated from the Poisson model of the traffic, due to the intrinsic dependencies among the packets of real network flows. The SLCorr detector, on the other hand, is the optimum detector for correlated network flows, which also results in reasonable detection performance for Poisson-modeled network flows. Comparing the results for the three different traffic types (Tables 1, 2, and 3), we observe that the ACTV and PASSV schemes perform the worst for the SSH traffic (TCP port 22); we explain this by the fact that SSH flows are more correlated compared to HTTPS and SMTP flows, as they are based on the typing behaviors of the human entities. Another general observation from the simulations is that the detection performance improves as the number of packets, $N$, increases.

In the second set of experiments, we run the simulated detection schemes to measure the false negative error rates. Again, we use the detection thresholds derived through the analysis in previous sections. In each simulation of the SLCorr and ACTV schemes, the candidate network flow is watermarked using the RAINBOW scheme (Section 3) and then a network delay is randomly selected and applied to that flow from a large pool of network delays measured over the Planetlab infrastructure [16] (the average standard deviation of the network delay is around 10ms). Likewise, for the PASSV simulations the candidate network flow is delayed similarly to simulate the network interference. The delayed flow is then correlated with the original flow (non-delayed, and non-watermarked) using each of the detection schemes. Tables 4, 5, and 6 show the false negative of the experiments for the three different detection schemes, evaluated for three different TCP ports. For the watermark detection schemes of SLCorr and ACTV the experiments are repeated for four different values of the watermark amplitude, i.e., $a = 10ms, 15ms, 20ms, 30ms$. Also, all of the simulations are run for different values of the watermark length, $N$. Results show that by choosing reasonable parameters for the RAINBOW watermark, the SLCorr and ACTV detection schemes result in very small false negative rates, comparable to those of the passive detection. Again, we see that increasing $N$ improves the detection performance.

In the third set of experiments, we evaluate the false positive error rate of the three detection schemes over highly correlated network flows. More specifically, we use flow traces corresponding to web browsing activities of human entities that target the *same* destination websites at different times and from different network locations[2]. Table 7 shows the false positive error rates

2. The traces are generated and provided to us by Xun Gong from UIUC

## TABLE 1

False positive rate of different detection schemes for port 443 network flows. Each experiment is run for 10000 different pairs of flows.

| N | Scheme | Detection metric | | | False Positive |
|---|--------|-----|-----|-----|------|
| | | Min | Avg | Max | |
| 25 | SLCorr | -0.005 | -0.0031 | 0.00012 | 0.0068 |
| | ACTV | -457.385 | -37.8203 | 14.1698 | 0.0151 |
| | PASSV | -245.249 | -35.6167 | 2.8426 | 0.0054 |
| 50 | SLCorr | -0.005 | -0.0039 | 0.0012 | 0.0002 |
| | ACTV | -503.655 | -36.8637 | 3.9970 | 0.0159 |
| | PASSV | -567.917 | -45.5303 | 2.8297 | 0.0009 |
| 100 | SLCorr | -0.005 | -0.0042 | -0.0004 | 0 |
| | ACTV | -515.555 | -33.2478 | -2.2095 | 0 |
| | PASSV | -555.857 | -44.0783 | 2.9567 | 0.0023 |
| 200 | SLCorr | -0.005 | -0.0042 | -2.5E-5 | 0 |
| | ACTV | -608.838 | -33.5721 | 0.9735 | 0.0005 |
| | PASSV | -559.164 | -43.2514 | 2.9535 | 0.0018 |

## TABLE 2

False positive rate of different detection schemes for port 25 network flows. Each experiment is run for 10000 different pairs of flows.

| N | Scheme | Detection metric | | | False Positive |
|---|--------|-----|-----|-----|------|
| | | Min | Avg | Max | |
| 25 | SLCorr | -0.005 | -0.0039 | 0.0018 | 0.0008 |
| | ACTV | -461.182 | -50.3404 | 6.1398 | 0.0003 |
| | PASSV | -364.275 | -49.6125 | 1.8952 | 0.003 |
| 50 | SLCorr | -0.005 | -0.0042 | 0.0004 | 0 |
| | ACTV | -359.413 | -35.2567 | -0.3314 | 0 |
| | PASSV | -364.652 | -53.7937 | 1.5171 | 0.0015 |
| 100 | SLCorr | -0.005 | -0.0037 | -0.0007 | 0 |
| | ACTV | -352.581 | -31.3738 | 0.0420 | 0.0001 |
| | PASSV | -368.304 | -55.4709 | 1.4271 | 0.0013 |
| 200 | SLCorr | -0.005 | -0.0041 | -0.0014 | 0 |
| | ACTV | -190.366 | -29.6399 | -1.2917 | 0 |
| | PASSV | -375.012 | -56.3069 | 1.3936 | 0.0012 |

## TABLE 3

False positive rate of different detection schemes for port 22 network flows. Each experiment is run for 10000 different pairs of flows.

| N | Scheme | Detection metric | | | False Positive |
|---|--------|-----|-----|-----|------|
| | | Min | Avg | Max | |
| 25 | SLCorr | -0.005 | -0.0029 | 0.0026 | 0.0024 |
| | ACTV | -495.125 | -18.3825 | 6.8506 | 0.0269 |
| | PASSV | -88.1381 | -8.7786 | 3.3239 | 0.1031 |
| 50 | SLCorr | -0.005 | -0.0038 | 0.0011 | 0.0001 |
| | ACTV | -628.45 | -20.1249 | 4.5654 | 0.0144 |
| | PASSV | -80.5081 | -9.3516 | 3.3204 | 0.0879 |
| 100 | SLCorr | -0.005 | -0.0037 | 0.0005 | 0 |
| | ACTV | -522.241 | -23.434 | 2.8119 | 0.0142 |
| | PASSV | -101.337 | -9.8241 | 3.3202 | 0.0861 |
| 200 | SLCorr | -0.005 | -0.0039 | 1.67E-5 | 0 |
| | ACTV | -487.594 | -26.357 | 4.7264 | 0.0212 |
| | PASSV | -104.547 | -9.7138 | 3.3195 | 0.0896 |

## TABLE 4

False negative rate of different detection schemes for port 443 network flows. Each experiment is run for 10000 different pairs of flows.

| N | Scheme | False Negative | | | |
|---|--------|------|------|------|------|
| | | 10 ms | 15 ms | 20 ms | 30 ms |
| 25 | SLCorr | 0.039 | 0.005 | 0.0004 | 0.0003 |
| | ACTV | 1E-04 | 1E-04 | 0 | 0.0004 |
| | PASSV | 0.0002 | | | |
| 50 | SLCorr | 0.0137 | 0.0004 | 0 | 0 |
| | ACTV | 0 | 0 | 0 | 0 |
| | PASSV | 0 | | | |
| 100 | SLCorr | 0.0028 | 0 | 0 | 0 |
| | ACTV | 0 | 0 | 0 | 0 |
| | PASSV | 0 | | | |
| 200 | SLCorr | 0.000977 | 0 | 0 | 0 |
| | ACTV | 0 | 0 | 0 | 0 |
| | PASSV | 0 | | | |

## TABLE 5

False negative rate of different detection schemes for port 25 network flows. Each experiment is run for 10000 different pairs of flows.

| N | Scheme | False Negative | | | |
|---|--------|------|------|------|------|
| | | 10 ms | 15 ms | 20 ms | 30 ms |
| 25 | SLCorr | 0.0346 | 0.0035 | 0.0007 | 0 |
| | ACTV | 0.0003 | 0.0002 | 0.0004 | 0.0002 |
| | PASSV | 0.0001 | | | |
| 50 | SLCorr | 0.0154 | 0.0005 | 0.0003 | 0 |
| | ACTV | 0 | 0 | 0 | 0.0006 |
| | PASSV | 0 | | | |
| 100 | SLCorr | 0.002636 | 0 | 0 | 0 |
| | ACTV | 0 | 0 | 0 | 0 |
| | PASSV | 0 | | | |
| 200 | SLCorr | 0 | 0 | 0 | 0 |
| | ACTV | 0 | 0 | 0 | 0 |
| | PASSV | 0 | | | |

## TABLE 6

False negative rate of different detection schemes for port 22 network flows. Each experiment is run for 10000 different pairs of flows.

| N | Scheme | False Negative | | | |
|---|--------|------|------|------|------|
| | | 10 ms | 15 ms | 20 ms | 30 ms |
| 25 | SLCorr | 0.028879 | 0.001775 | 0 | 0 |
| | ACTV | 0.0003 | 0 | 0.00062 | 0.005727 |
| | PASSV | 0.0002 | | | |
| 50 | SLCorr | 0.009671 | 0 | 0 | 0 |
| | ACTV | 0 | 0 | 0 | 0 |
| | PASSV | 0 | | | |
| 100 | SLCorr | 0 | 0 | 0 | 0 |
| | ACTV | 0 | 0 | 0 | 0 |
| | PASSV | 0 | | | |
| 200 | SLCorr | 0 | 0 | 0 | 0 |
| | ACTV | 0 | 0 | 0 | 0 |
| | PASSV | 0 | | | |

for different detection schemes for different websites and for different values of $N$ (each simulation is averaged over 100 runs). As can be seen, in most of the case, the ACTV and PASSV detection schemes result in very high false positive rates, while the SLCorr scheme results in *no false positive error in all of the cases*. This confirms what we expect intuitively: *the PASSV and ACTV scheme are optimum passive and active detection schemes for independent network traffic models, but they perform poorly as the network flows get more correlated*. The SLCorr scheme, however, is the optimum detection scheme for correlated network flows, *and* it also performs good enough in the case of independent network flows.

## 6 Conclusions

In this paper, we introduce the first non-blind active traffic analysis scheme, RAINBOW. Using the tools from the detection and estimation theory, we find the optimum passive and (non-blind) active traffic analysis schemes for different types of the network flows. We show that, for different traffic models, the optimum active detectors outperform the optimum passive detectors. This advantage is more significant for the more correlated network traffic, e.g., the web browsing traffic. Considering the fact that both passive and non-blind active approaches of traffic analysis are constrained by similar scalability issues, this finding motivated the use of non-blind active approaches over the passive approaches.

## Appendix A
## Chernoff bounds

*Lemma 1 (Chernoff bound for signal detection):*
Consider the following binary hypothesis testing for signal detection:

$$\begin{cases} H_0: & y_i \sim p_{0,i}(y_i) \quad i = 1, ..., n \\ H_1: & y_i \sim p_{1,i}(y_i) \quad i = 1, ..., n \end{cases} \tag{100}$$

For this hypothesis testing consider a detection scheme with rule:

$$T(\mathbf{y}) \underset{H_0}{\overset{H_1}{\gtrless}} \eta$$

such that $T(\mathbf{y}) = \sum_{i=1}^{n} T_i(y_i)$.

We are interested in finding the false positive rate $P_{FP} = Pr\{T(y) > \eta\}$ and the false negative rate $P_{FN} = Pr\{T(y) < \eta\}$ of this detector in different cases. We have that:

a) General case:

$$P_{FP} \leq e^{-(\eta s - \mu_0^T(s))} \qquad (s > 0) \tag{101}$$

$$P_{FN} \leq e^{-(\eta s - \mu_1^T(s))} \qquad (s < 0) \tag{102}$$

where $\mu_k^T(s)$ is the cumulant generating function (CGF) of $T(\cdot)$ under hypothesis $H_k$.

b) Independent $T_i(\cdot)$'s: We have that:

$$\mu_k^T(s) = \sum_{i=1}^{n} \mu_k^{T_i}(s)$$

where $k$ corresponds to hypothesis $H_k$. This results in the error rates to be:

$$P_{FP} \leq \prod_{i=1}^{n} e^{-(s\eta/n - \mu_0^{T_i}(s))} (\forall s > 0) \tag{103}$$

$$P_{FN} \leq \prod_{i=1}^{n} e^{-(s\eta/n - \mu_1^{T_i}(s))} (\forall s < 0) \tag{104}$$

For $T_i(y_i) = \ln[\frac{p_{1,i}(y_i)}{p_{0,i}(y_i)}]$, this reduces to

$$P_{FP} \leq \prod_{i=1}^{n} e^{-(s\eta/n - \mu_{0,i}(s))} \tag{105}$$

$$P_{FN} \leq \prod_{i=1}^{n} e^{-n((s-1)\eta/n - \mu_{0,i}(s))} \tag{106}$$

$$(0 < s < 1)$$

where:

$$\mu_{0,i}(s) = \ln \int p_{0,i}^{1-s}(y) p_{1,i}^s(y) dy \tag{107}$$

c) i.i.d. $T_i(\cdot)$'s: For any $i$ and $j$ we have that $\mu_k^{T_i}(s) = \mu_k^{T_j}(s) = \mu_k^{T_1}(s)$, which reduces the false error rates to:

$$P_{FP} \leq e^{-n(s\eta/n - \mu_0^{T_1}(s))} \qquad (\forall s > 0) \tag{108}$$

$$P_{FN} \leq e^{-n(s\eta/n - \mu_1^{T_1}(s))} \qquad (\forall s < 0) \tag{109}$$

For $T_i(y_i) = \ln[\frac{p_{1,i}(y_i)}{p_{0,i}(y_i)}]$, this reduces to

$$P_{FP} \leq e^{-n(s\eta/n - \mu_0(s))} \tag{110}$$

$$P_{FN} \leq e^{-n((s-1)\eta/n - \mu_0(s))} \tag{111}$$

$$(0 < s < 1)$$

where:

$$\mu_0(s) = \ln \int p_{0,1}^{1-s}(y) p_{1,1}^s(y) dy \tag{112}$$

## Appendix B
## Summation of random variables

*Lemma 2 (Summation of two Laplacian random variables):*
Suppose that we have two independent random variables distributed according to Laplacian distribution as $X \sim Lap(0, 1/\alpha)$ and $Y \sim Lap(0, 1/\beta)$ where $\alpha \neq \beta$. The PDF function of the summation of these random variables, $Z = X + Y$, is given by:

$$f_Z(z) = \frac{\alpha\beta}{2(\alpha^2 - \beta^2)} \left( \alpha e^{-\beta|z|} - \beta e^{-\alpha|z|} \right) \tag{113}$$

If $\alpha = \beta$ then:

$$f_Z(z) = \frac{\alpha^2}{4} \left( |z| + \frac{1}{\alpha} \right) e^{-\alpha|z|} \tag{114}$$

*Proof:* Using the convolution of PDFs:

$$f_Z(z) = (f_X * f_Y)(z)$$

□

*Lemma 3 (Summation of Laplacian and Exponential r.v.s):*
Suppose that $X \sim Exp(\lambda)$, and $Y \sim Lap(0, b)$. The

TABLE 7

False positive error rate of different detection schemes for network flows generated by browsing the same websites.

| Website | N=25 | | | N=50 | | | N=100 | | |
|---|---|---|---|---|---|---|---|---|---|
| | SLCorr | ACTV | PASSV | SLCorr | ACTV | PASSV | SLCorr | ACTV | PASSV |
| baidu.com | 0 | 0.08 | 0.29 | 0 | 0.12 | 0.07 | 0 | 0.12 | 0.08 |
| blogger.com | 0 | 0.56 | 0.97 | 0 | 0.89 | 0.63 | 0 | 0.34 | 1 |
| facebook.com | 0 | 0.95 | 0.91 | 0 | 0.9 | 0.97 | 0 | 0.59 | 0.96 |
| live.com | 0 | 0.81 | 1 | 0 | 0.33 | 1 | 0 | 0.08 | 0.38 |
| wikipedia.org | 0 | 0.44 | 0.94 | 0 | 0.44 | 0.44 | 0 | 0.39 | 0.46 |
| yahoo.co.jp | 0 | 0.08 | 0.66 | 0 | 0.03 | 0.33 | 0 | 0 | 0.05 |
| yahoo.com | 0 | 1 | 1 | 0 | 0.02 | 1 | 0 | 0 | 0.23 |
| yandex.com | 0 | 0.11 | 0.89 | 0 | 0.02 | 0.08 | 0 | 0 | 0.02 |

random variable $Z = X + Y$ has the following distribution:

$$f_Z(z) = \begin{cases} \frac{\lambda}{2(\lambda b - 1)} e^{-\frac{z}{b}} + \frac{\lambda}{1 - \lambda^2 b^2} e^{-\lambda z} & z \geq 0 \\ \frac{\lambda}{2(\lambda b + 1)} e^{\frac{z}{b}} & z < 0 \end{cases} \quad (115)$$

Also, for a fixed integer $m$, the random variable $T = Z - m$ has the PDF:

$$f_T(t) = f_Z(t + m)$$

We abbreviate this as:

$$f^{EL}(t, m, \lambda, b) = f_T(t)$$

*Proof:* Using the convolution of PDFs:

$$f_Z(z) = (f_X * f_Y)(z)$$

$\square$